\documentclass{aa}
\usepackage{graphics}
\usepackage{times}
\newcommand{\gc}{\hbox{$\gamma$~Cas }}
\newcommand{\gce}{\hbox{$\gamma$~Cas}}

\newcommand{\cv}{\v{c}}

\newcommand{\ii}{\'{\i}}

\newcommand{\rv}{\v{r}}
\newcommand{\sv}{\v{s}}

\newcommand{\Sv}{\v{S}}

\newcommand{\aaa}{A \&~A }

\newcommand{\aas}{A\&~AS }

\newcommand{\an}{Astron. Nachr. }
\newcommand{\apj}{ApJ }

\newcommand{\apss}{Ap\&~SS }

\newcommand{\bac}{Bull. Astron. Inst. Czechosl. }

\newcommand{\hob}{Hvar Obs.Bull. }
\newcommand{\iau}{IAU Circ. No.}

\newcommand{\mn}{MNRAS }

\newcommand{\pasj}{PASJ }
\newcommand{\pasp}{PASP }

\newcommand{\ssr}{Space Sci. Rev. }

\newcommand{\p}{$\pm$}

\newcommand{\D}{$^{\rm d}\!\!.$}

\newcommand{\kms}{km~s$^{-1}$ }
\newcommand{\ks}{km~s$^{-1}$}

\newcommand{\tef}{$T_{\rm eff}$ }

\newcommand{\ms}{M$_{\odot}$}
\newcommand{\rs}{R$_{\odot}$}
\newcommand{\ha}{H$\alpha$ }

\newcommand{\he}{He~{\sc I}~6678 }
\newcommand{\hea}{He~{\sc I}~6678}

\newcommand{\AAA}{\accent'27A}
\newcommand{\Am}{\AA~mm$^{-1}$ }

\begin{document}
 \thesaurus{08.02.4, 08.05.2, 08.09.2(\gce)}
 \title{Properties and nature of Be stars}
 \subtitle{XX. Binary nature and orbital elements of \gc}
  \author{P.~Harmanec\inst{1,2}\and P.~Habuda\inst{3}\and
  S.~\v{S}tefl\inst{2}\and P.~Hadrava\inst{2}\and
  D.~Kor\v{c}\'akov\'a\inst{4,2}\and P.~Koubsk\'y\inst{2}\and
  J.~Krti\v{c}ka\inst{4,2}\and J.~Kub\'at\inst{2}\and
  P.~\v{S}koda\inst{2}\and M.~\v{S}lechta\inst{2}\and M.~Wolf\inst{1}
  }
  \offprints{P. Harmanec: hec@sunstel.asu.cas.cz}
  \institute {
   Astronomical Institute of the Charles University,
   V Hole\sv ovi\cv k\'ach~2, CZ-180~00~Praha 8, Czech Republic
\and
   Astronomical Institute of the Academy of Sciences,
   CZ-251~65~Ond\rv ejov, Czech Republic
\and
   Faculty of Mathematics and Physics of the Charles University,
   Ke~Karlovu~3, CZ-121~16~Praha 2, Czech Republic
\and
   Department of Theoretical Physics and Astrophysics,
   Faculty of Science, Masaryk University,
   CZ-611 37~Brno, Czech Republic
}

\authorrunning{P. Harmanec et al.}
\titlerunning{Spectroscopic binary orbit of \gc}

\date{Received September 19, 2000, accepted November 27, 2000}

\maketitle

\begin{abstract}
An analysis of accurate radial velocities (RVs) of the Be star \gc
from 295 Reticon spectrograms secured between October 1993 and May 2000
allowed us to prewhiten the RVs for the long-term changes
and to obtain the first orbital RV curve of this star. The orbital period
is 203\D59 and the orbit has an eccentricity of 0.26. The orbital motion
is detectable even in the published velocities, based on photographic
spectra. This implies that \gc is a primary component of a spectroscopic
binary. The secondary has a mass of about 1~\ms,
appropriate for a white dwarf or a neutron star, but it could also be
a normal late-type dwarf. The ultimate solution of the dispute whether
the observed X-ray emission is associated with the secondary or with
the primary will need further dedicated studies.
\end{abstract}
\section{Introduction}
The Be star \gc~(27~Cas, HD~5394, HR~264, ADS~782A) is a member of
a visual multiple system -- see also \cite{gonc} who discovered a new
companion, closer than ADS~782B. It is the very first Be star known,
discovered by \cite{secchi}. It exhibits spectral
and light variations on several distinct time scales. History of its
pronounced long-term spectral and light variations was summarized, e.g., by
\cite{doaz83} or \cite{telt94}. It underwent two consecutive shell phases
in 1935-36 and 1939-40, followed by a relatively short phase when it
appeared as a normal B star. When the Balmer emission lines are present,
they exhibit cyclic long-term $V/R$ and radial-velocity
changes. In 1976, \cite{jer} and \cite{mason} reported that \gc
is the optical counterpart of the X-ray source MX0053+60.
However, the attempts to detect the orbital motion of \gc (on the
assumption that \gc is a binary with a compact companion), led to
negative results - see \cite{cow} and \cite{jarad}.
There is also evidence of rapid spectral changes of \gc in the form of
line-asymmetry and RV variations on a characteristic time scale of
0\D7 -- see \cite{john} and \cite{jarad} -- and in the form of travelling
sub-features returning every about 0\D8 to the line centre  -- see
\cite{ninkov, yang, hora} and \cite{smith95}.
Later, \cite{smith1} found simultaneous but anti-correlated
UV and X-ray light changes with a period of 1\D123
which they identified with the rotational period of \gce.
Note, however, that \cite{tow3}, searching for such a period only in a close
neighbourhood of the 1\D123 period, found a significantly different
period of 1\D15655\p0\D00012 from the Hipparcos $H_p$ band photometry.
The complex nature of rapid line-profile changes of the UV Si~IV lines
was demonstrated by \cite{smith2}.

There are currently two competing interpretations of the nature of
the observed X-ray emission: one is the accretion of the wind from \gc onto
a putative white-dwarf companion and the other one is that it originates
from some physical processes in the outer atmosphere of \gc itself.
Arguments for and against these two hypotheses are best summarized
in recent studies by \cite{kubo} and \cite{rob2000}.

Realizing that there had been no attempt to search for the orbital
motion of \gc in electronic spectra, we have been collecting
Reticon spectrograms of \gc since 1993. Here, we report the first results.

\begin{figure}
 \resizebox{\hsize}{!}{\includegraphics{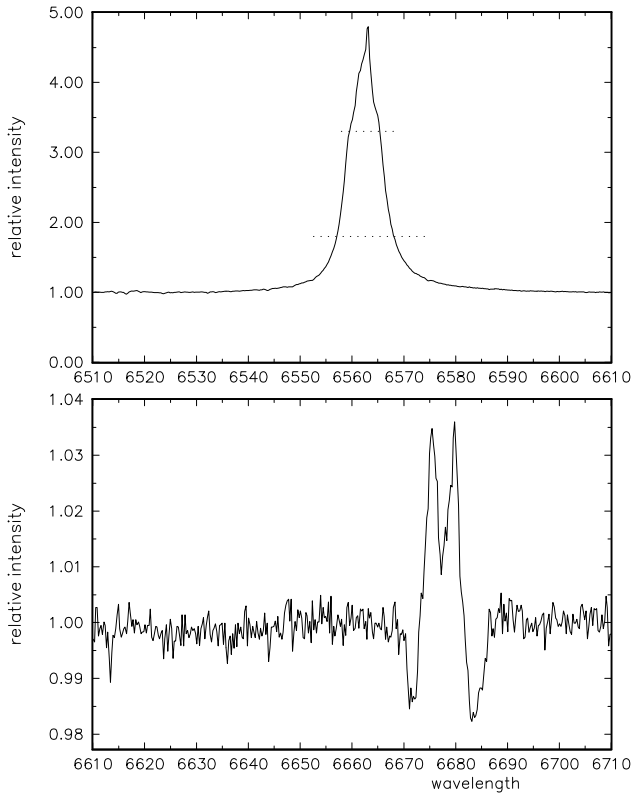}}
\caption[ ]{The \ha and He~I~6678 line profiles from an Ond\v{r}ejov
Reticon spectrum of \gc obtained on HJD~2449310.1969.
Note the different intensity scale at the two panels}\label{sp}
\end{figure}

\section{Observations and reductions}
All 295 spectrograms used here have been obtained in the coud\'e focus of the
Ond\rv ejov 2-m reflector with a Reticon 1872RF detector. They cover the
wavelength region from 6280 to 6720~\AAA\ and have a linear dispersion
of 17.1~\Am (4~pixels per \AA). A standard reduction of these spectrograms
was carried out using program SPEFO, developed by Dr.~J.~Horn -- see
\cite{sef0} and \cite{spefo}. The zero point of RV scale was corrected
through the use of reliable telluric lines. The two strongest lines seen,
\ha and \hea, from one Reticon spectrum are shown in Fig.~\ref{sp}.
We measured RVs of \ha (emission wings) and of \he line
(absorption and emission wings and the shell absorption core)
interactively, comparing the direct and reverse images of
the profiles. For the \ha profile, we set on the symmetric steep portions of
the emission profile (having the relative intensities between the two
dotted lines in Fig.~\ref{sp}), taking care to avoid disturbances due to
telluric lines. Although the \ha profile is complicated, we found that
independent measurements by two of us led to an excellent reproduction
of the results, usually within 2 \ks. All our RV measurements will be
published in a follow-up study, devoted to the long-term changes of \gce.

\section{Analysis of RV changes}
\begin{figure}
 \resizebox{\hsize}{!}{\includegraphics{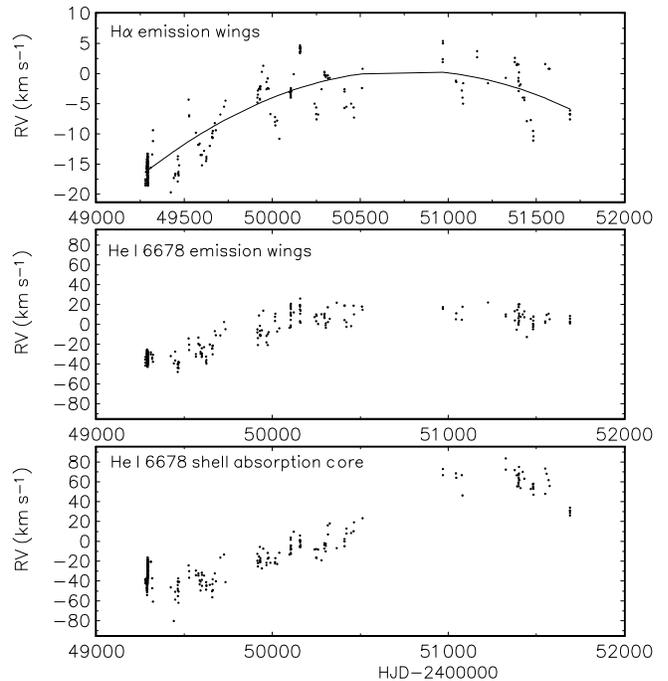}}
\caption[ ]{RVs of several spectral features plotted vs. time:
The solid line for \ha emission RVs shows the spline function used
to the removal of the long-term changes}\label{time}
\end{figure}

Figure~\ref{time} is a plot of RVs of several measured lines vs. time.
Large RV changes, typical for long-term variations of Be stars, are clearly
visible, the He~I shell core having a larger RV amplitude than the emission
wings. Our data confirm further lengthening of the cycles, demonstrated
already by \cite{telt94}. In addition to these variations, there is
a pattern of apparently {\it regular} RV changes of the \ha emission wings
on a time scale of about 200 days. This prompted us
to prewhiten the RVs of the \ha emission for the long-term changes using
spline smoothing after \cite{von69} and \cite{von77} -- see the solid
line in Fig.~\ref{time} -- and to analyze the RV residuals for periodicity.
It immediately turned out that a very well-defined periodic variation with
a period of about 203-204~d is present. (The same is also true about the
RV of the He~I emission and He shell core which give residual 203-d RV curves
in phase with the \ha emission but with a larger scatter because the He line
is weak.) It was demonstrated, e.g., by \cite{hrvoje} for the primary of
the double-lined Be binary $\varphi$~Per (and also found for several other
known Be binaries) that the steep symmetric parts of the \ha emission wings
may be used to the detection of the orbital motion of a Be star.
We therefore adopt a working hypothesis that the RV changes of the \ha
emission with a period of 203~d reflect the orbital motion of \gce.

\begin{figure}
 \resizebox{\hsize}{!}{\includegraphics{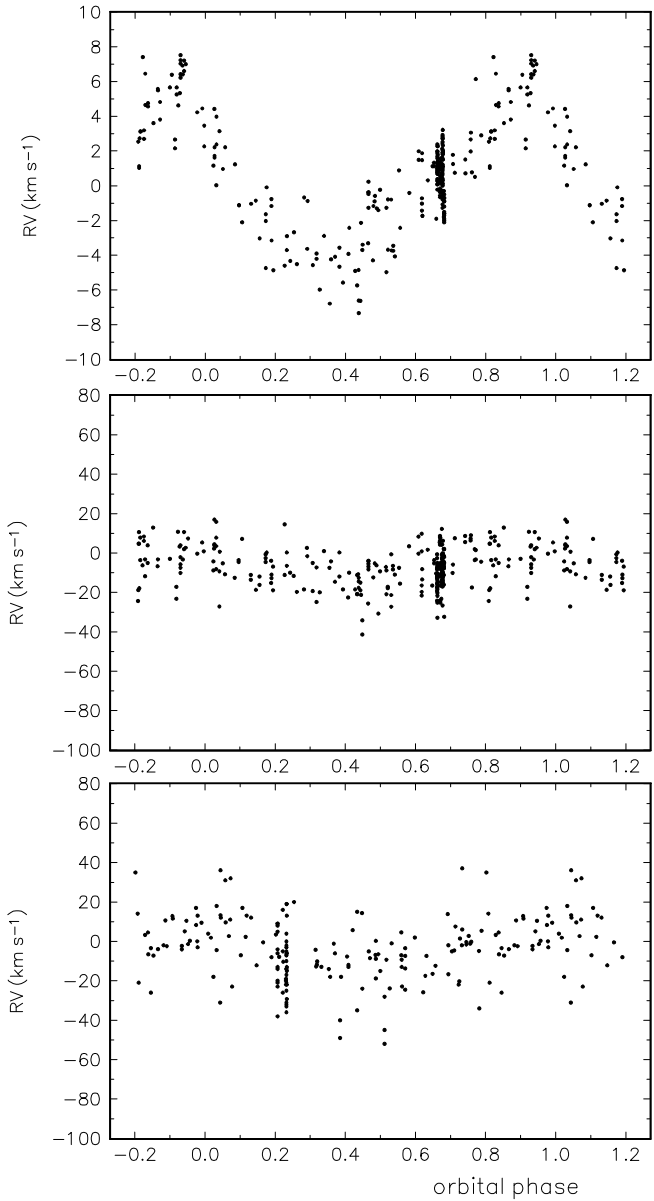}}
\caption[ ]{Orbital RVs curves of \gc plotted with the ephemeris derived
from the \ha emission RVs:
$T_{\rm peri.}$ = HJD~2450578.7 + 203\D59$\times E$.
From top to bottom: \ha emission wings,
broad absorption wings of the \he line
from Ond\v{r}ejov spectra, and absorption RVs from the photographic spectra
published by \cite{cow} and \cite{jarad}
}\label{ph}
\end{figure}

\begin{figure}
 \resizebox{\hsize}{!}{\includegraphics{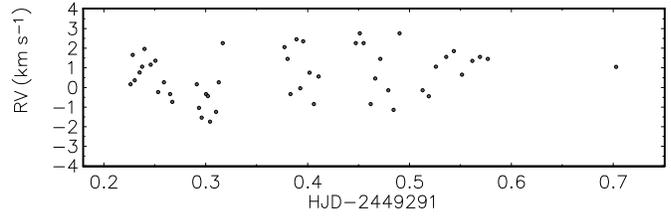}}
\caption[ ]{RV residuals of the \ha emission velocities from the longest
night series of observations plotted vs. time. No trend indicative of rapid
changes is seen}\label{noc}
\end{figure}

\section{\gc as a spectroscopic binary}
Using the program FOTEL, developed by \cite{fotel}, we derived the orbital
elements using the \ha emission RVs prewhitened for the long-term changes.
They are given in the second column of Tab.~\ref{orbit} and the
phase diagram is shown in the upper panel of Fig.~\ref{ph}. The RV
curve is defined remarkably well, especially when one considers its small
semi-amplitude. To characterize the measuring accuracy and
to demonstrate that the variations occur indeed on the time
scale of 203\D59 and not on a time scale close to 1~day,
we plot, in Fig.~\ref{noc}, the \ha emission RVs vs. time for our longest
night series. Only some scatter but no systematic trend
can be seen there. We also derived orbital solutions for our \he
absorption-wing RVs and for published RVs. There is a large scatter around
the phase curve and one would be hesitant to accept this result without
the evidence from the \ha emission RVs but the curves are in phase and
all orbital elements mutually agree within the limits of their errors --
cf. Fig.~\ref{ph}. A significant part of the scatter may be due to the
sub-features moving accross the line profiles which affect the blue and
red wing differently at different times. It is also encouraging to see
that a free convergency of the period for RVs from photographic spectra
(which span 16000~days) leads to a value very similar to that derived
from the Reticon data only. RVs of the wings of He~I emission and of
its shell absorption core, prewhitened for the long-term changes, also
follow the 203.6-d period in phase with \ha emission and give
semiamplitudes of 7.7 \p 1.2 \kms and 7.5 \p 1.2 \ks, respectively.
A consistent detection of a truly periodic RV variation in several spectral
lines, originating in different parts of the photosphere and the Be
envelope, constitutes a rather solid proof of binary nature of \gce.

\begin{table}
\caption{Orbital solutions for the Ond\v{r}ejov Reticon velocities
of \ha emission wings prewhitened for long-term changes, \he
absorption wings and absorption RVs from photographic spectra.
Separate systemic velocities were derived for the RVs by
\cite{cow} (C) and \cite{jarad} (J). All epochs are in HJD-2400000,
$K_1$, $\gamma$ and rms error per 1 observation are in \ks,
No. is the number of RVs}
\label{orbit}
\begin{flushleft}
\begin{tabular}{lllllll}
\hline\noalign{\smallskip}
Element                & \ha emission & \he abs. &photogr. \\
\noalign{\smallskip}\hline\noalign{\smallskip}
$P$ (d)                &203\D59\p0.29 &203\D59 fixed &203.62\p0.15    \\
$T_{\rm peri.}$        &50578.7\p4.2  &50576\p16     &38391\p18       \\
$T_{\rm upper\, c.}$   &50592.8       &50599.8       &38401           \\
$T_{\rm lower\, c.}$   &50513.9       &50528.8       &38319           \\
$e$                    &0.260 \p 0.035&0.260 fixed   &0.260 fixed     \\
$\omega\,(^\circ)$     &47.9\p8.0     &23\p27        &57\p37          \\
$K_1$                  &4.68\p0.25    &7.0\p1.5      &11.0\p2.0       \\
$\gamma$               & --           &-7.38\p0.64   &-3.3\p1.0 C     \\
                       &              &              &-8.4\p2.7 J     \\
rms                    &1.455         &8.946         &14.13           \\
No.                    &272           &280           &169             \\
\noalign{\smallskip}\hline
\end{tabular}
\end{flushleft}
\end{table}

\section{Probable basic physical properties of \gc}

There is a rather large uncertainty concerning the true effective
temperature of \gc since the star has strong emission lines all the time
since fifties. Perhaps a more probable is the high limit of log~\tef\ = 4.51
derived from line-profile modelling by \cite{john} and from spectrophotometry
fits by \cite{goraya}. For the often quoted spectral class B0.5 it would be
log~\tef\ = 4.46. According to mass vs. log~\tef\ calibration by
\cite{masses}, this range corresponds to masses of 18 and 13~\ms,
respectively, for the primary of \gce. Another uncertainty lies in
the true value of the amplitude of the orbital motion of the primary.
While the RV curve based on emission RVs certainly defines well the
orbital period, its amplitude may or may not reflect the orbital motion
properly. However, it agrees quite well with the RV amplitude
of the \he absorption wings.
Note that the interferometric resolution of the Be disk by \cite{quir}
implies an inclination $i>44^\circ$ if the disk lies in the equatorial
plane of the binary and that also \cite{john} arrived at an inclination
near about 50$^\circ$.
Using the above range of primary masses, semi-amplitudes between
4.7 and 7 \kms and a plausible range of orbital inclinations
45$^\circ$ to 90$^\circ$, one finds the following properties of the system:
the mass of the secondary between 0.7 and 1.9~\ms\ and a separation
of the binary components between 350 and 400~\rs\ (250 -- 300~\rs\
at periastron). The secondary could, therefore, be the long-expected
hot compact object but also a late-type star of much lower luminosity
than the primary.

\section{Discussion}

\begin{figure}
 \resizebox{\hsize}{!}{\includegraphics{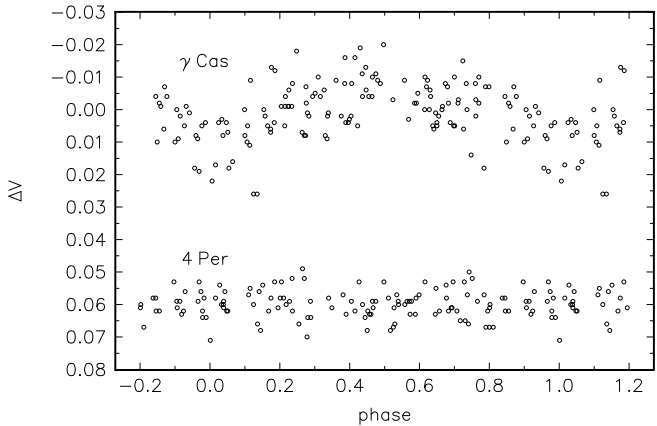}}
\caption[ ]{A plot of Hipparcos photometry, prewhitened for a linear
trend, vs. phase of the best-fit short period of 1\D48700. Phases are
calculated from the epoch of minimum light, HJD~2448350.838.
To illustrate the noise of Hipparcos photometry, the original $H_p$
photometry of a constant 5-mag. B star 4~Per, suitably shifted in magnitude,
is also shown}\label{hipv}
\end{figure}

The latest arguments by \cite{rob2000} against the origin of the X-ray
emission from an accretion onto a white-dwarf companion appear quite
convincing. However, the understanding of the physical processes involved
is still far from complete. We mention two possibly relevant items:
(i) We found that \gc is a binary with a 203\D59 period and that its orbit
is eccentric ($e=0.26$), similarly as for other Be+X binaries.
Other known Be binaries usually have circular orbits. The varying
distance between the components should be seen to affect the putative
accretion onto the secondary. A search for possible modulation of
the X-ray flux with the phase of the orbital period is, therefore, highly
desirable.
(ii) Considering the importance of the 1\D12 period for the alternative
hypothesis, we re-analyzed good (flag = 0) Hipparcos $H_p$ photometry
of \gce, prewhitened for a linear trend, over a wider range of possible
short periods than \cite{tow3} did: between 0\D6 and 5\D0. The best-fit
period is 1\D48700\p0\D00013, with minimum light at
HJD~2448350.838\p0.027 -- see Fig.~\ref{hipv} which shows the superiority
of this period to the period found by \cite{tow3}. The new period is very
significantly different from 1\D12. If real, it could be either a corotation
period at some level or a low-order mode of pulsation.

It is, therefore, desirable to carry out further tests
of both competing hypotheses.

\begin{acknowledgements}
We devote this study to the memory of our late colleague and friend,
Dr.~Ji\v{r}\ii\ Horn, who started this program with us, obtained some
of the spectra and developed the program SPEFO, used for the spectral
reductions. Our sincere thanks are due to Mr. J.~Havelka who helped with
the initial reductions of the spectra. We also thank to Drs.~V.~\v{S}imon
and T.~Raja who obtained some of the spectra and put them at our disposal.
The use of the bibliography from the CDS at Strasbourg is gratefully
acknowledged. Comments of the referee, Dr. M.A.~Smith, helped us to
improve the paper. This study was supported from the grants 436 TSE 113/19
(Academy of Sciences of the Czech Republic and Deutsche Forschungsgemeinschaft)
and 202/97/0326 of the Grant Agency of the Czech Republic to S.\v{S},
A3003805 of the Grant Agency of the Academy of Sciences of the Czech Republic
to P.Had. and also from the research plan K1-003-601/4. Research of P.Har. and
M.W. was supported from the research plan J13/98: 113200004.
\end{acknowledgements}

\end{document}